\documentclass[12pt]{article}
\usepackage[affil-it]{authblk}
\usepackage{amssymb,physics,slashed,graphicx}
\begin{document}
	
	\title{\bf  Black-Hole-Like Saturons in Gross-Neveu}
	\author{Gia Dvali}
	\author{Otari Sakhelashvili}

	\affil{Max-Planck-Institut f\"ur Physik, F\"ohringer Ring 6, 80805 M\"unchen, Germany}
	\affil{Arnold Sommerfeld Center, Ludwig-Maximilians-Universit\"at, Theresienstra{\ss}e 37, 80333 M\"unchen, Germany}
	
	\maketitle

\begin{abstract}
It has been argued recently that objects of maximal microstate entropy permitted by unitarity, the so-called saturons, have properties similar to black holes. We demonstrate the existence of such objects  in Gross-Neveu model.  From the large-$N$ scaling of $S$-matrix, we deduce the connection between the entropy of the bound-state and the unitarity of scattering. We observe that upon saturation of unitarity, the bound state exhibits a remarkable  correspondence with a black hole. The scaling of its entropy is identical to Bekenstein-Hawking entropy.  The saturon decays via Hawking's thermal rate of temperature given by the inverse size. The information retrieval time from the Gross-Neveu saturon is isomorphic to Page's time.  Our observations indicate that black hole properties are exhibited by saturated states in simple calculable models. 

\end{abstract}
	\pagebreak

\section*{Introduction}

Black holes are considered to be mysterious due to the properties concerning their entropy, the thermal evaporation and the time-scale of information retrieval. However, it has been argued recently  \cite{Dvali:2020wqi} (for a summary,  see \cite{Dvali:2021jto}) that these properties are not specific to black holes or to gravity. Rather,  they are the universal features of objects that have maximal microstate entropy compatible with unitarity.This  proposal heavily relies on the evidence produced in two previous articles \cite{Dvali:2019ulr,Dvali:2019jjw} and is  based on several field theoretic examples. The connection goes as follows.
 
 In a theory with a quantum coupling $\alpha$, the entropy attained by any self-sustained object of size $R$ is bounded from above by \cite{Dvali:2020wqi}, 
\begin{equation}
	S\leq\frac{1}{\alpha} \,, 
	\label{unba}
\end{equation}
where the coupling $\alpha$ has to be evaluated at the scale $1/R$.  A violation of the above bound, would lead to a violation of unitarity. This, of course, is impossible. Instead, if pushed beyond (\ref{unba}), the theory has to resist and undergo a change of the regime.  

It has been shown that the saturation of the bound (\ref{unba}) is correlated with a breakdown of loop expansion, as well as, with the saturation of unitarity by certain scattering amplitudes. An intuitive way of understanding this connection is the following. An object with unbounded entropy $S$, would lead to an unlimited growth of the cross section due to an exponentially increasing number of the degenerate final states.

On more quantitative side, near the saturation point, a typical scaling of the relevant cross section has the form \cite{Dvali:2020wqi},  
   \begin{equation}
	\sigma  \sim  e^{-\frac{1}{\alpha}  + S} \,. 
	\label{2n_cross}
\end{equation}
This  makes the bound  (\ref{unba}) rather transparent. It has been argued \cite{Dvali:2020wqi} that objects saturating this bound, so-called ``saturons", behave similar to black holes. 
  
The basis for this connection becomes clear by realizing that Bekenstein-Hawking entropy \cite{Bekenstein:1973ur, Hawking:1975vcx} of a black hole of radius $R$, 
 \begin{equation} \label{BHE}
   S_{BH}    \sim \frac{\mathcal Area}{G_{gr}}  \,,  
\end{equation} 
  can be written as \cite{ Dvali:2020wqi,Dvali:2011aa}, 
  \begin{equation}
	S \sim  \frac{1}{\alpha_{gr}} \,, 
	\label{BH}
\end{equation}
where  $\alpha_{gr} = G_{gr} /R^{d-2}$ is the quantum gravitational coupling in $d$ space-time dimensions, evaluated at the scale of momentum transfer $1/R$.  The quantity ${\mathcal Area} \sim R^{d-2}$ and $G_{gr}$ stands for a $d$-dimensional Newton's constant. 
 
Remarkably, the black holes and gravity turn out not to be exceptional. The entropy of a generic saturated state obeys a very similar area-law \cite{Dvali:2020wqi, Dvali:2019ulr,Dvali:2019jjw}. For a generic saturon, the role of Newton's constant $G_{gr}$ is played by the coupling $G_{Gold}$ of the  Nambu-Goldstone boson of spontaneously broken Poincare symmetry.  Correspondingly, the maximal entropy  (\ref{unba}) can be presented in the following Goldstone form
 \cite{Dvali:2020wqi,Dvali:2019ulr}\,, 
  \begin{equation} \label{SEA}
   S   \leqslant \frac{\mathcal Area}{G_{Gold}} \,.    
\end{equation}  
This form holds for saturons in arbitrary dimensions. The coupling $G_{Gold}$ is well defined due to an universal character of spontaneous breaking of Poincare symmetry by any saturated state. For a black hole, the Goldstone comes from graviton, and correspondingly, $G_{Gold} = G_{gr}$. But the relation (\ref{SEA}) is generic.

Thus, uncovering a hidden correspondence between black holes and other saturated objects, requires a formulation of entropy  in terms of universal parameters, such as $\alpha$ and $G_{Gold}$.  
  
It was argued \cite{Dvali:2020wqi} that saturons and black holes share other key features. Such are, the Hawking type evaporation, as well as, the minimal time-scale of the information retrieval. For a generic saturated object, this minimal time is given by, 
   \begin{equation}
	t_{min}  = \frac{R}{\alpha} = SR \,. 
	\label{time}
\end{equation}
No relevant information can be obtained on any shorter time-scale. For $\alpha = \alpha_{gr}$ this expression reproduces the so-called Page's time \cite{Page:1993wv} for a black hole.

The above similarities between black holes and generic saturons were  demonstrated on several explicit examples  \cite{Dvali:2020wqi,Dvali:2019ulr,Dvali:2019jjw, Bubble}. These include solitons, instantons, baryons and other field configurations, such as the vacuum bubbles. 

 The fact that the black-hole-like properties follow from saturation, creates a supporting evidence for the description of a black hole as saturated bound state of gravitons \cite{Dvali:2011aa}. It has been argued  recently \cite{Dvali:2021ooc} that the color glass condensate state in ordinary QCD \cite{Gelis:2010nm} saturates the bounds (\ref{unba}) and (\ref{SEA}) and exhibits a correspondence with the black hole state in gravity, described as the saturated state of gravitons of \cite{Dvali:2011aa}.    
 
 In is important to say that the correspondence between black holes and other saturons \cite{Dvali:2020wqi},  does not imply the correspondence  between gravity and the respective theories. Rather, it refers to a correspondence between the specific states in different theories. The connection emerges  due to the universal relation between the maximal entropy (\ref{unba}) and unitarity. 
 
 The purpose of the present paper is to demonstrate the above connection in a theory for which the $S$-matrix is calculable while simultaneously the spectrum of bound states is known.  An interesting  candidate  of this sort is the Gross-Neveu model \cite{Gross:1974jv}. This theory has all the required features. In particular, it possesses  the bound-states of high degeneracy for which we can explicitly test the ideas about saturation.

We shall demonstrate that the bound state of maximal degeneracy in Gross-Neveu model is a saturon. Due to saturation,  the Gross-Neveu  bound state exhibits the black hole like properties. The correspondence includes all the key physical characteristics, such as, the entropy given by (\ref{unba}) and (\ref{SEA}); a Hawking-like evaporation, with the effective temperature given by the inverse size of the saturon;  the minimal time-scale (\ref{time}), required for information retrieval. 

In this way, the Gross-Neveu model captures and explains the physical origin of these ``mysterious" properties,  in terms of the universal  phenomenon of saturation.

\section{The Gross-Neveu model}

 In a well-known work by Gross and Neveu \cite{Gross:1974jv}, a 2-dimensional quantum field theory model was discussed (for the review, see, \cite{Manohar:1998xv,Coleman:1980nk}). The model is integrable in large $N$ limit. The chiral symmetry is spontaneously broken and the bound states are known  \cite{Dashen:1975xh}.

 The model consists of $N$ Dirac $\psi$ fermions in 2 dimensions. Alternatively,  we can work in the bases of $2N$ Majorana fermions with $SO(2N)$ flavor symmetry. The Lagrangian has the following form,
\begin{equation}
\mathcal{L}=i\bar{\psi}\slashed\partial\psi+\frac{\alpha}2\left(\bar\psi\psi\right)^2 \,,
\label{GN}
\end{equation}
where $\alpha$ is a dimensionless coupling constant of the theory. The contractions of flavor and space-time indexes are obvious and not shown explicitly. The theory has a discrete chiral symmetry, which in Dirac notations acts as $\psi\rightarrow\gamma_5\psi$. The above theory can be written in the equivalent form,
\begin{equation}
	\mathcal{L}=i\bar{\psi}\slashed\partial\psi-\frac1 {2\alpha} \sigma^2 +\sigma\bar\psi\psi \,,
	\label{sigma_form}
\end{equation}
where $\sigma$ is an auxiliary scalar field. It is a singlet under the flavor group and transforms under the discrete chiral symmetry as $\sigma\rightarrow -\sigma$. 

 We adopt the normalization for the $\sigma$ field (\ref{sigma_form}) used in \cite{Manohar:1998xv}. Of course, the physical quantities, such as masses of particles and the vacuum expectation value of the $\bar\psi\psi$ operator are independent of this normalization. 

Analysis of the model is usually done in the large $N$ limit, where the `t Hooft-like coupling \cite{tHooft:1973alw} is defined as
\begin{equation}
 \lambda=\alpha N \,. 
 \label{thooft}
\end{equation}
We shall be working in the 't Hooft limit, in which $N \rightarrow \infty$, whereas the collective coupling $\lambda$ is kept fixed. In this case,  the full effective potential for $\sigma$ is known \cite{Gross:1974jv} and has the form, 
\begin{equation}
	V=\frac{N}{2\lambda}\sigma^2+\frac{N}{4\pi}\sigma^2\left(\log(\frac{\sigma^2}{\mu^2})-1\right),
	\label{eff_pot}
\end{equation}
where $\mu$ is the renormalization scale.  The above potential has two minima at $\sigma_0=\pm\mu e^{-\frac{2\pi}{\lambda}}$. This implies a spontaneous breaking of the chiral symmetry. The result is exact for  infinite $N$. 

The beta function corresponding to the coupling $\lambda$ has the following form
\begin{equation}
	\beta(\lambda)=-\frac{\lambda}{2\pi},
\end{equation}
 which can be derived from the effective potential (\ref{eff_pot}). The sign of the beta function implies that the model is asymptotically free. This statement is also exact, since the effective potential is exact in large $N$. 

Due to the spontaneous breaking of symmetry, the elementary fermions get the following masses,
\begin{equation}
	m_f=\sigma_0\,.
	\label{fermionM}
\end{equation}
We shall treat the fermions as the building blocks of the asymptotic $S$-matrix states. Due to the fermionic nature of the constituents, the wave functions form the totally antisymmetric representations of the flavor group. The number of states $n_{st}$ in such a tensor multiplet of rank $n$ is given by a binomial coefficient, 
\begin{equation}
n_{st}=\frac{2N!}{n!(2N-n)!}.
\label{micro_GN}
\end{equation}
The semi-classical study of the theory shows that it has bound states. They form multiplets with the above degeneracy \cite{Dashen:1975xh}. The mass spectrum has the following form
\begin{equation}
	M_n \, = \, m_f \frac{2N}{\pi}\sin(\frac{n}{N}\frac\pi2),
	\label{spectrum}
\end{equation}
where $n \, <  \, N$. It is clear that for large $N$, the case $n=1$ reproduces the mass of an elementary fermion. 

All the above information makes the Gross-Neveu model a perfect laboratory for testing the entropy bounds (\ref{unba}) and (\ref{SEA}) and studying their relation with unitarity.  

\section{The entropy of bound states} 

Following \cite{Dvali:2020wqi, Dvali:2019ulr,Dvali:2019jjw}, we shall now apply the concept of entropy to the bound states. Due to degeneracy, the bound states of Gross-Neveu model carry the following entropy, 
 \begin{equation} \label{SSSA}
  S = \ln (n_{st}) \,,
 \end{equation} 
  where  the number of microstates $n_{st}$ is given by (\ref{micro_GN}).  This quantity reaches its maximum for $n = N$.  Due to large $N$, we can use Stirling approximation, which gives, 
  \begin{equation} \label{SSS}
  S =  2N \ln (2)\,.  
 \end{equation} 
 This expression is valid up to relative corrections of order $\ln(N)/N$, which are negligible.  The equation (\ref{SSS}) describes the limiting entropy of the bound state in Gross-Neveu model.   From (\ref{spectrum}), the mass of this bound state is given  by 
  \begin{equation}
	M_N \, = \, \frac{2N}{\pi}m_f\,. 
	\label{MMM}
\end{equation}
Thus, both the mass and the entropy of the bound state scale as $N$ in units of the fermion mass $m_f$. We also note that the size of the bound state is set by the Compton wavelength of the fermion, 
\begin{equation} \label{RRR}
R\sim \frac{1}{m_f} \,.
\end{equation} 

\subsection{Correspondence with black holes} 

 We shall now argue that characteristics of this object are very similar to a black hole, when parameters are translated into their gravitational counterparts.
  
For this let us notice that the bound state saturates the unitarity entropy bound (\ref{unba}), when the 't Hooft coupling (\ref{thooft}) becomes order one.  This behaviour is rather characteristic to saturated states in large-$N$ theories, as discussed in \cite{Dvali:2020wqi, Dvali:2019ulr,Dvali:2019jjw}. In this limit,  the entropy of the bound state becomes, 
     \begin{equation} \label{Slimit}
  S \sim  \frac{1}{\alpha} \sim N \,. 
 \end{equation} 
 This is the same expression as the entropy of a black hole  (\ref{BH})  expressed in terms of the gravitational coupling $\alpha_{gr}$.  Thus, we discover that  for the critical 't Hooft coupling (which, as we will show, saturates unitarity), there is a correspondence between the entropies of a black hole (\ref{BH}) and of the heaviest Gross-Neveu bound state. The correspondence is exact under  the exchange,  $\alpha  \rightarrow  \alpha_{gr}$.

\subsection{Area-law in Gross-Neveu} 

When written in terms of the Newton's  constant, $G_{gr}$, the black hole entropy assumes a well-known area form  (\ref{BHE}). In order to understand the meaning of the area-scaling, we must appreciate that it originates from the form of the graviton coupling (\ref{BH}) at the point of saturation. Thus, the  black hole entropy satisfies the combined relation.    
\begin{equation} \label{FGNA}
   S_{BH}   \sim \frac{\mathcal Area}{G_{gr}} \sim  \frac{1}{\alpha_{gr}} \,.  
\end{equation} 
  
One may wonder, what is the analog of $G_{gr}$ for Gross-Neveu saturon.   As explained in  \cite{Dvali:2020wqi}, the role of the Newton's coupling,  in case of a generic saturon, is played by the coupling $G_{Gold}$ of the Goldstone boson of spontaneously broken Poincare symmetry.  In terms of $G_{Gold}$, the entropy bound (\ref{unba})  takes the form (\ref{SEA}). In each case, the coupling $G_{Gold}$ is unambiguously defined. This is because any object of non-zero entropy, inevitably breaks a part of Poincare symmetry spontaneously. 

 In particular, for a generic self-sustained saturated state of $N$ quanta, of characteristic wavelength $R$,  the coupling $G_{Gold}$ is  universally given by, 
 \begin{equation} \label{Fscale}
   G_{Gold}   \sim  \frac{R^{d-2}}{N} \sim \frac{R^{d-2}}{S} \,.
\end{equation} 
Correspondingly, the form of entropy, 
 \begin{equation} \label{FGN}
   S   \sim \frac{\mathcal Area}{G_{Gold}} \sim  \frac{1}{\alpha} \,,  
\end{equation} 
is universal. This is also true in Gross-Neveu. The only peculiarity is that in this case, since $d=2$,  the coupling of the Poincare Goldstone,  is dimensionless $G_{Gold}  = 1/\alpha = N$. This is no obstacle for the validity of  (\ref{FGN}), since the quantity ``${\mathcal Area}$",  which for $d>2$ is interpreted as the surface  area,  continues to be well defined for $d=2$.  It is simply equal to a dimensionless number, ${\mathcal Area} \equiv R^{0} \sim 1 \neq 0$. Therefore, all relations in (\ref{FGN}) are valid for the Gross-Neveu saturon. 
   
The correspondence between a black hole and a generic saturon,  goes through the invariant notions that are independent of the space-time dimensionality and other peculiarities of the theory. Such trans-theory invariants are the couplings, $\alpha$ and $G_{Gold}$, evaluated at the scale of the object.  It is this coupling that constrains the entropy of the object via (\ref{unba}). When the bound is saturated, the object becomes a saturon. The correspondence with a black hole then follows.  
  
  \subsection{Connection with Bekenstein} 
  Another indication of the deep connection is that the bound state, just like a black hole, simultaneously with 
  (\ref{unba}) and (\ref{SEA}),  saturates  the Bekenstein bound on entropy. The Bekenstein bound says that the entropy of an object of  mass $M_N$ and radius $R$ is bounded by \cite{Bekenstein:1980jp},   
 \begin{equation}\label{bek} 
S =2\pi M_NR\,. 
\end{equation}
Taking into account the size (\ref{RRR}) and the mass (\ref{MMM}), this expression takes the form 
 \begin{equation}\label{BN} 
S \sim N\,,  
\end{equation}
which coincides with (\ref{SSS}). Thus, the  Bekenstein enropy bound on the Gross-Neveu saturon coincides with the microstate entropy of the maximally degenerated state.

\subsection{Connection with unitarity} 

Can the the entropy of the maximally degenerate state (\ref{SSS}) violate the bound (\ref{unba})? Since the entropy scales as $N$, the violation of the bound (\ref{unba}) would require  $N \gg 1/\alpha$, or equivalently, 
\begin{equation} \label{LLL}
\lambda  \gg 1\,.  
\end{equation}
Thus, the  violation of (\ref{unba}),  would require the increase of the 't Hooft coupling $\lambda$ beyond the critical point. This is not possible due to the following reasons \cite{Dvali:2020wqi,Dvali:2019ulr,Dvali:2019jjw}.  

First, for (\ref{LLL}),  the loop expansion breaks down. Secondly, the saturation of the entropy bound (\ref{unba}) is correlated with the saturation of unitarity by scattering amplitudes. Both connections indicate that the theory must prevent the violation of the entropy bound, by changing the regime, whenever  $\lambda$ is increased above the saturation point. The Gross-Neveu model allows us to explicitly check these  statements.

\section{Saturation of scattering amplitudes}

Following the general criteria of  \cite{Dvali:2020wqi}, we shall now establish the correlation between the entropy bound and unitarity in Gross-Neveu model.

The basic point is simple.  The connection goes through the 't Hooft  coupling $\lambda$.  As (\ref{Slimit}) indicates, the saturation of the entropy bound (\ref{unba}) coincides with the regime in which $\lambda$ is becoming order one. Of course, $\lambda$ has to be understood as the running coupling evaluated at the scale $m_f$. This scale sets the size of the bound state according to (\ref{RRR}). 

This gives the first indication that the entropy cannot exceed the bound (\ref{unba}).  No further increase of $\lambda$, without jeopardising the properties of the bound-state, is possible. If we try to push the entropy (\ref{SSS}) above (\ref{unba}), by increasing $\lambda$, we shall not succeed.  Instead the theory will change the regime.

The fact that $\lambda \sim 1$ represents a critical point, is visible from number of factors. The first clear sign is the breakdown of loop expansion, in series of $\lambda$. The fact that loop expansion is controlled by $\lambda$, is demonstrated by the following diagram (using the form of Lagrangian 
(\ref{sigma_form})),    
\begin{equation}
		\includegraphics[scale=0.5]{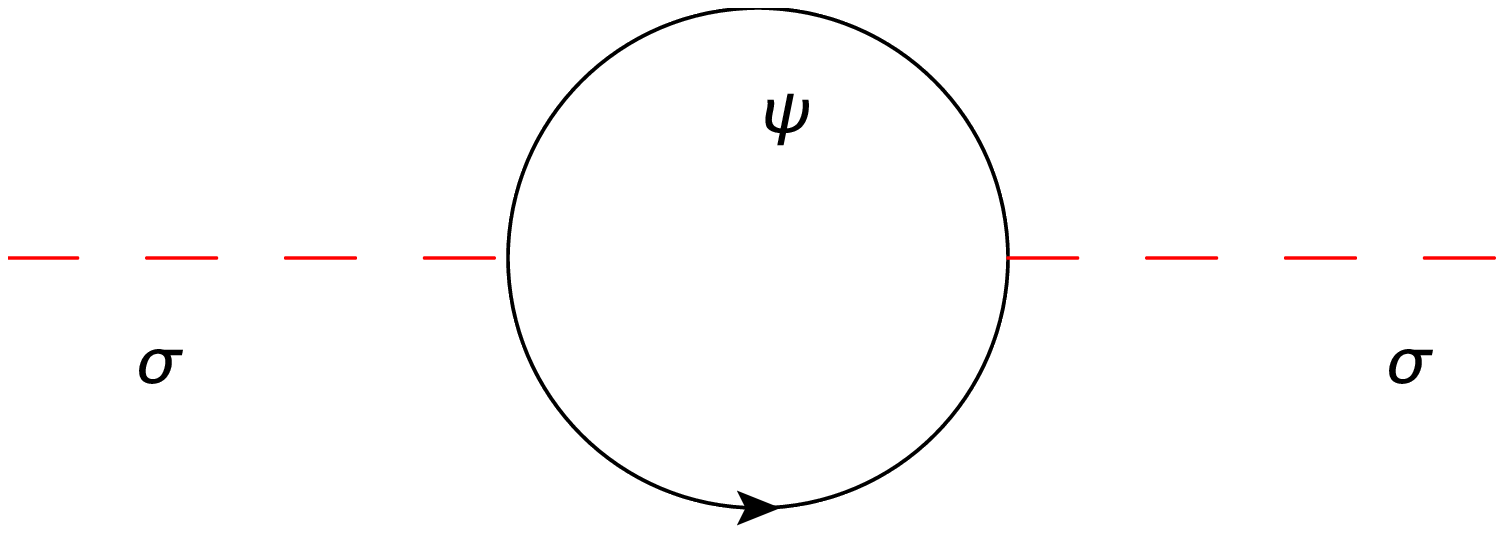}\centering,
\end{equation}
where dashed and solid lines represent $\sigma$ scalar  and $\psi$ fermions respectively.  Since there are  $\sim N$ fermion flavors running in the loop, this diagram is of order $\lambda=N\alpha$. When this quantity hits the critical value, $\lambda \sim 1$,  the loop expansion breaks down and series must be resummed. 
 
Alternatively, we can see a similar breakdown from the corrections to the fermion self-energy (using the Lagrangian (\ref{GN})). For example, the two-loop contribution of the following type,
 \begin{equation}
		\includegraphics[scale=0.5]{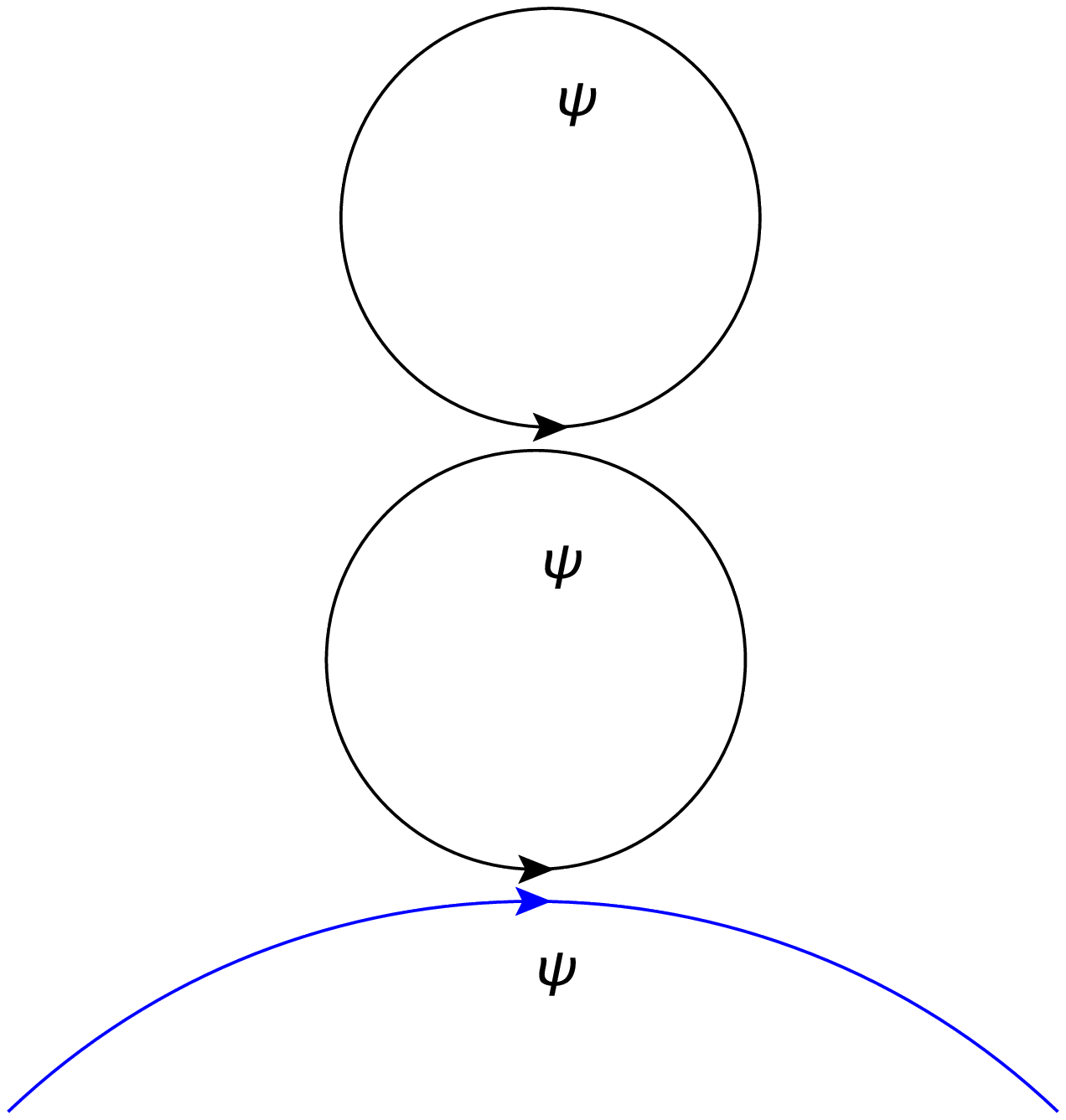}\centering,
 \end{equation}
is of order $\lambda^2$. Each extra bubble brings an additional power of $\lambda$.  The expansion breaks down around $\lambda \sim 1$.  
 
Another indication that the excessive entropy must trigger the regime-change, comes from tree-level scattering amplitudes. For $\lambda \sim 1$,  various $S$-matrix elements become order one. Let us first illustrate this for $2 \rightarrow 2$ transition. The large-$N$ expression for $S$-matrix in this case was derived in \cite{Zamolodchikov:1977hh}. 

We shall look for a transition from an $SO(2N)$-invariant two-particle state with the opposite momenta. We shall compute the transition element from this state into a similar state. That is, the initial and final states have the form, 
 \begin{equation}
 \ket{in} = \ket{f} = \frac{1}{\sqrt{2N}} \sum_i \ket{i,p}\ket{i,-p} \,,  
 \end{equation} 
where $\ket{i,p}$ describes a one-particle state of a Majorana fermion of flavor $i$ and momentum $p$. The transition amplitude at large $N$ takes the following form, 
\begin{equation}
{\rm Amp}_{in \rightarrow f}  
= \lambda  \,. 
 \end{equation} 
For simplicity, we wrote the expression up to an overall kinematic factor. The  exact form of it can be reconstructed, e.g., from  \cite{Zamolodchikov:1977hh}.  For our purposes, it suffices to know  that in 't Hooft limit this factor is independent of $N$. We see that the $S$-matrix element tends to saturate unitarity for $\lambda \sim 1$.  Of course, $\lambda$ cannot be made arbitrarily large,  as its critical value sets the scale of the dimensional transmutation $m_f$.  Correspondingly, there is no way to push the entropy of any bound state beyond  (\ref{unba}). 
 
This resistance is reminiscent of the entropy of a baryon  in $SU(N)$ QCD in four space-time dimensions \cite{Dvali:2020wqi, Dvali:2019jjw}. The entropy of the baryon, similarly to the entropy  of  the Gross-Neveu bound state, saturates (\ref{unba}) when the number of quark flavors is $\sim N$. A further increase of entropy of the baryon, would require a substantial increase of the flavor symmetry relative to color. This is not possible, without  abolishing the asymptotic freedom of the theory. 

Next we move to multi-particle amplitudes. This connection is especially interesting, since the saturation of unitarity is observed precisely when the number of final state quanta becomes comparable to the number of constituent fermions in the maximal entropy bound-state.  This is an universal tendency also observed in theories of higher dimensionality \cite{Dvali:2020wqi}. Correspondingly, the saturation of unitarity by such diagrams can be interpreted as the unitarization of the scattering process by creation of a high entropy ``classical" state, a saturon. A nice thing about the Gross-Neveu model is that semi-classically  the spectrum is known exactly.  Due to this,  the candidate state for saturation of unitarity is naturally identified as the bound state of the highest  entropy.  As already discussed, it consists of $N$ fermions. As a consistency check, we shall correlate its degeneracy with the scaling of a $2N$-fermion scattering process. 

We shall work in Dirac basis. In the theory (\ref{GN})  we prepare a state in which we have the equal numbers $n = \bar{n}$ of fermions ($\psi$)  and anti-fermions ($\bar\psi$).  We shall denote such a state by
\begin{equation}
	\ket{n, \bar{n}}\,.
\end{equation}
The fact that the fermions can assume different flavors, allows to give them the similar characteristic momenta. 

Let us consider an $S$-matrix process of a transition from the  vacuum $\ket{\Omega}$ to the state of $n$ particle-anti-particle  pairs,
\begin{equation}
	\bra{n, \bar{n}}\hat{S} \ket{\Omega} \,.
\end{equation}
 This process obviously is not an on-shell one, but its amplitude allows to catch the scaling with respect to the particle number $n$.  A more ``realistic" on-shell processes of the type $1+\bar1\rightarrow n+\bar{n}$, in which the state $\ket{n, \bar{n}}$ is created from two initial quanta of high center of mass energy, can be analysed in the same manner.  

We compute the amplitude at the lowest order in the perturbation series in $\alpha$, 
\begin{equation}
{\rm Amp} \, \propto \, \bra{n \bar{n}}\frac{1}{n!}\frac{1}{2^n}\alpha^nT\left(:\bar\psi\psi\bar\psi\psi:\right)^n\ket{0} \,.
\end{equation}
 Notice, although in 't Hooft limit $\alpha \rightarrow 0$, the transition can be meaningfully analysed due to the high  degeneracy of the final state. As already said, our main interest is to extract the scaling with respect to $n$.  We therefore shall not derive Green functions. 

Following the usual routine and collecting all the factors, the resulting amplitude of the process scales in the following way,
\begin{equation}
{Amp} \, \propto \, \frac{1}{n!}\frac{1}{2^n}\alpha^n\times \left(n!\right)^2 2^n\, =\, \alpha^n n! \,.
\end{equation}
The corresponding cross section scales as, 
\begin{equation}
	\sigma \,
	\propto \, 
(\alpha^n n!)^2 \,. 
	\label{cross1}
\end{equation}
 Notice that for any given  value of  $\alpha$,  this cross section blows up for large $n$,  due to the factorial growth. This growth is typical for $n$-particle processes \cite{Voloshin:1992rr}.  It signals the breakdown of $\alpha$-expansion for sufficiently large $n$.  Beyond this, the series must be re-summed. 

We shall work at  the point of optimal truncation, for which the above expression gives the most reliable approximation for the process. This point is given by,  
\begin{equation} \label{optimal} 
 n \, = \,  \frac{1}{\alpha} \,.
\end{equation}
The error is $ \sim 1/n$. 

At the point of optimal truncation, the cross section scales in the following way, 
\begin{equation} \label{cross2} 
	\sigma \,
	\propto  \, \left(n!n^{-n}\right)^2 \sim e^{-2n}\, = \, e^{-\frac{2}{\alpha}} \,
\end{equation}
where the Stirling approximation holds for large $n$, up to unimportant factors. This represents a particular example of the general scaling (\ref{2n_cross}). 

 We wish to relate the above computation to the production of the Gross-Neveu bound-state of the maximal entropy. For this we must take $n=N$.  This already gives the first glimpse of the connection between the maximal entropy and the saturation of unitarity. Indeed, notice that, with this value of $n$, the point of optimal truncation (\ref{optimal}) gives $\lambda = 1$. As already discussed, this value of 't Hooft coupling is critical.  Beyond it, the theory must change the regime. 

The total cross section of creation of the maximal entropy state, is obtained by the summation over all degenerate microstates. This is equivalent to multiplication of (\ref{cross2}) by the degeneracy factor, ${\rm e}^{S}$.  We obtain, 
\begin{equation} \label{crossT} 
\sigma_{total}\, = \, \sum_{micro  state}\sigma \propto e^{-2N +  S}=e^{-\frac{2}{\alpha} + S }\,. 
\end{equation}
Let us have a closer look at the above expression.
 
First, this expression reproduces the bound (\ref{unba}) in the context of the Gross-Neveu theory. We see that the maximal entropy, permitted by unitarity, assumed by the state of $N$ quanta, is  $S \sim 1/\alpha \sim N$. It also indicates that the entropy of the bound-state (\ref{SSS})  is strikingly close to achieving the saturation.  In particular, it has the right scaling with $N$. 
 
Notice, the expression (\ref{SSS}) carries an extra factor of $\ln(2)$,  as compared to the critical value $S= 2N$ required for a complete saturation of (\ref{crossT}). This mismatch however does not imply that the saturation of the cross section by pair-creation of maximal entropy bound-states is incomplete. The reason is that, at the level of the present analysis, there is an uncertainty, which can be interpreted in two ways.  
 
First,  the expression (\ref{crossT}) is reliable when $S$ approaches the saturation point from below. That is, from the region of weak $\lambda$.  Very close to the saturation point, the higher corrections in $\lambda$  must be resummed.  This cannot change the fact that the saturation takes place somewhere in $\lambda \sim 1$ domain. Correspondingly, the bound (\ref{unba}) is robust. However, it leaves an uncertainty in exact saturation of the process  by  $n=N$ bound states.

Another reasoning is that the discrepancy follows from our approximation of describing the bound state as the state of $N$ free fermions.  This approximation creates an error.  The magnitude can be estimated in the following way.  
 
The fermions in the bound state are off-shell as compared to their free counterparts. This is due to the interaction energy. Since we are working in 't Hooft limit,  $\alpha$ is vanishingly small. Correspondingly, interactions between the individual fermion pairs are infinitely weak. However, the collective effect is non-zero. The binding potential experienced by each fermion from the rest of $N$ fermions, is $ \sim N\alpha m_f \sim \lambda m_f$. 
 
For a weak 't Hooft coupling, $\lambda \ll 1$, this is a negligible correction to the fermion self-energy. Correspondingly, in this case,  the fermion can be regarded as free. However, we have seen that the saturation of the entropy bound happens  when $\lambda \sim 1$. In this case the collective interaction cannot be ignored. The resulting potential puts the fermion off-shell.
 
Due to this, there are good news and bad news. The bad news is that the approximation of a free fermion is no longer exact. That is, when approximating the production of bound states by the state of $2N$ free fermions, we are committing an error. 
 
The good news however is that the off-shellness is only of order one. In particular, the wavefunction overlap between a fermion in the bound state and its would-be free version, is order one, rather than vanishingly small. Correspondingly, the cross section of producing a true bound-state is expected to differ from the one of $2N$ free fermions by an exponential factor ${\rm e}^{c_{\lambda}2N}$. Here  $c_{\lambda}$ is an unknown number  that depends on the collective coupling and is $\sim 1$ for  the critical value $\lambda =1$.  Assuming that the maximal entropy bound state of Gross-Neveu theory is a saturon,  we estimate $c_{\lambda} = 1-\ln(2) \simeq 0.3$. 

Regardless the exact saturation, the tendency is rather transparent. We see that the theory resists to an unlimited growth of the entropy of any $n$-particle state by hitting the unitarity bound. This is clear both from  the breakdown of perturbation theory of loop expansion, as well as, from the saturation of unitarity by the scattering amplitudes.

Those are the signals how the theory protects itself from violating unitarity, thereby, restricting the degeneracy of the states. This restriction is manifested as the bound (\ref{unba}). In the present case this translates as the bound on $N$,   
\begin{equation}
 N  \lesssim  \frac{1}{\alpha} \,.
\end{equation}

We must note that the evident saturation of the scattering cross section by the maximal entropy bound states in Gross-Neveu model, is strikingly similar to saturation of $2 \rightarrow N$ graviton scattering by black holes \cite{Dvali:2014ila, Addazi:2016ksu}. In that analysis too, the cross section of the individual $N$-particle states is suppressed by the factor ${\rm e}^{-\frac{1}{\alpha_{gr}}}$, which is compensated by the entropy of the black hole.

\section{Time-scale of information retrieval} 
  
We have seen that the maximal degeneracy bound state in Gross-Neveu model,  saturates the same entropy  bounds as the black hole in gravity. Correspondingly, its entropy has exact same scaling properties in terms of parameters of the theory. We now wish to discuss, yet another similarity with the black hole:  The time of information retrieval.  
  
As argued in \cite{Dvali:2020wqi}, for arbitrary saturons, the minimal time-scale required for the start of the information retrieval is bounded from below by (\ref{time}). In particular, for a black hole this is a familiar expression.  However, the scale is not specific to gravity but rather comes from general features of saturation. We wish to show that the same applies to Gross-Neveu bound state.

The saturon of Gross-Neveu (as any other saturon) carries a maximal quantum information permitter by unitarity. The information is encoded in the flavor quantum numbers. Basically, the saturon represents an information pattern. The pattern can be read out, if we determine in which flavor state the saturon is.  This can be done in two ways. 
 
If saturon is unstable, one can simply wait and analyse the decay products. We shall consider such a possibility in the next chapter, when we discuss the analog of Hawking evaporation in Gross-Neveu. 
 
 The second option is to detect the flavor state of the saturon by performing a scattering experiment. That is,  an external probe can be scattered at the saturon, and the products of scattering can be analysed. Consider for example a scattering probe in form of a free fermion.  This forms a vector representation of $SO(2N)$ symmetry group. The interaction rate is suppressed by the coupling $\alpha^2$, but is enhanced by the number of fermions in the boundstate $N$.  The resulting scattering rate is, 
\begin{equation} \label{rateP}
 \Gamma \sim m_f \alpha^2 N \sim \frac{m_f}{N} \,, 
\end{equation}
 where the factor $m_f$ comes from the characteristic energy of the fermions and in the last expression we took into account the saturation condition $N \alpha \sim  1$.  The above expression sets the minimal time-scale of information retrieval, 
 \begin{equation} \label{TMIN}
 t_{min} \sim \frac{1}{\Gamma}  \sim \frac{N}{m_f}  \sim  S R\,.  
\end{equation}
 This reproduces the general expression (\ref{time}) for the minimal information retrieval time from any saturon. In particular, it agrees with the information retrieval time from a black hole, suggested by Page \cite{Page:1993wv}. Of course, in the present case the black hole radius is replaced by the size of the bound-state $R \sim 1/m_f$. 
 
We must note that (\ref{TMIN}) represents an absolute lower bound for the start of information retrieval. The actual time scale for the information read-out is much longer.

\section{Hawking Evaporation in Gross-Neveu}  
 
Another common feature that saturons share with black holes is their decay, which is strikingly similar to Hawking's thermal evaporation \cite{Dvali:2020wqi}.  We must make this point very clear. 
 
Just like black holes, saturons need not be unstable. For example, the saturon in the minimal Gross-Neveu model, which we have studied above, is stable, at least in semi-classical theory \footnote{In fact, the stability can be understood in terms of a so-called  ``memory burden" effect discussed in \cite{Dvali:2018xpy}. The essence of the phenomenon is that the quantum information carried by the system can stabilize it. Indeed, the information encoded in the flavor quantum numbers of the Gross-Neveu saturon, if released in form of elementary fermions, would cost more energy that the mass of the boundstate. Stabilization of solitons by memory burden is discussed in detail in \cite{Bubble}.}. This is due to the flavor symmetry. Likewise, if the additional conserved quantum numbers are introduced in a theory of gravity, some black holes  can become stable. 
 
However, similarity with Hawking evaporation emerges whenever saturons are unstable.  That is, if the theory obeys the saturation relation,  the saturons decay at the rate that is isomorphic to Hawking's thermal evaporation rate.  Originally, this mechanism of the emergent thermality was discussed in \cite{Dvali:2011aa}.
It has also been demonstrated on other examples of saturons, such as the maximal entropy vacuum bubbles \cite{Bubble}.

 We shall now arrive to the same effect in Gross-Neveu by an explicit calculation. In order to make the saturon unstable, we shall use a minimal extension of the type  discussed in the original article by Gross and Neveu. Namely, we shall introduce an additional massless pseudoscalar multiplet $\pi_{ij}$, transforming under the adjoint representation of $SO(2N)$ symmetry.  Here $i,j=1,2,...,2N$ are $SO(2N)$ indexes and $\pi_{ij}$ is an antisymmetric tensor. We shall refer to it as ``pion". 
 
 For convenience, we shall work in Majorana basis, in which the fermions are real spinors $\psi_j$ in the vector representation of $SO(2N)$. The coupling between the fermions and pions has the following form,
 \begin{equation} \label{pion}
   \frac{m_f}{f_{\pi}}\,  \pi_{ij} \, \bar{\psi_i} \gamma_5\psi_j \,,   
\end{equation}
where $f_{\pi}$ is the dimensionless pion decay constant. The saturation demands that it scales as $f_{\pi} \sim \sqrt{N}$.  The same scaling is demanded by the consistency of 't Hooft's planar limit \cite{Witten:1979kh}.

 The existence of the massless pion makes the saturon unstable. Let us estimate the decay rate.  The leading order contribution comes from the re-scattering of the constituent fermions, resulting in a single pion emission. The typical diagrams contributing to such a process are depicted in (\ref{fig2}).
 \begin{equation}
	\includegraphics[scale=0.4]{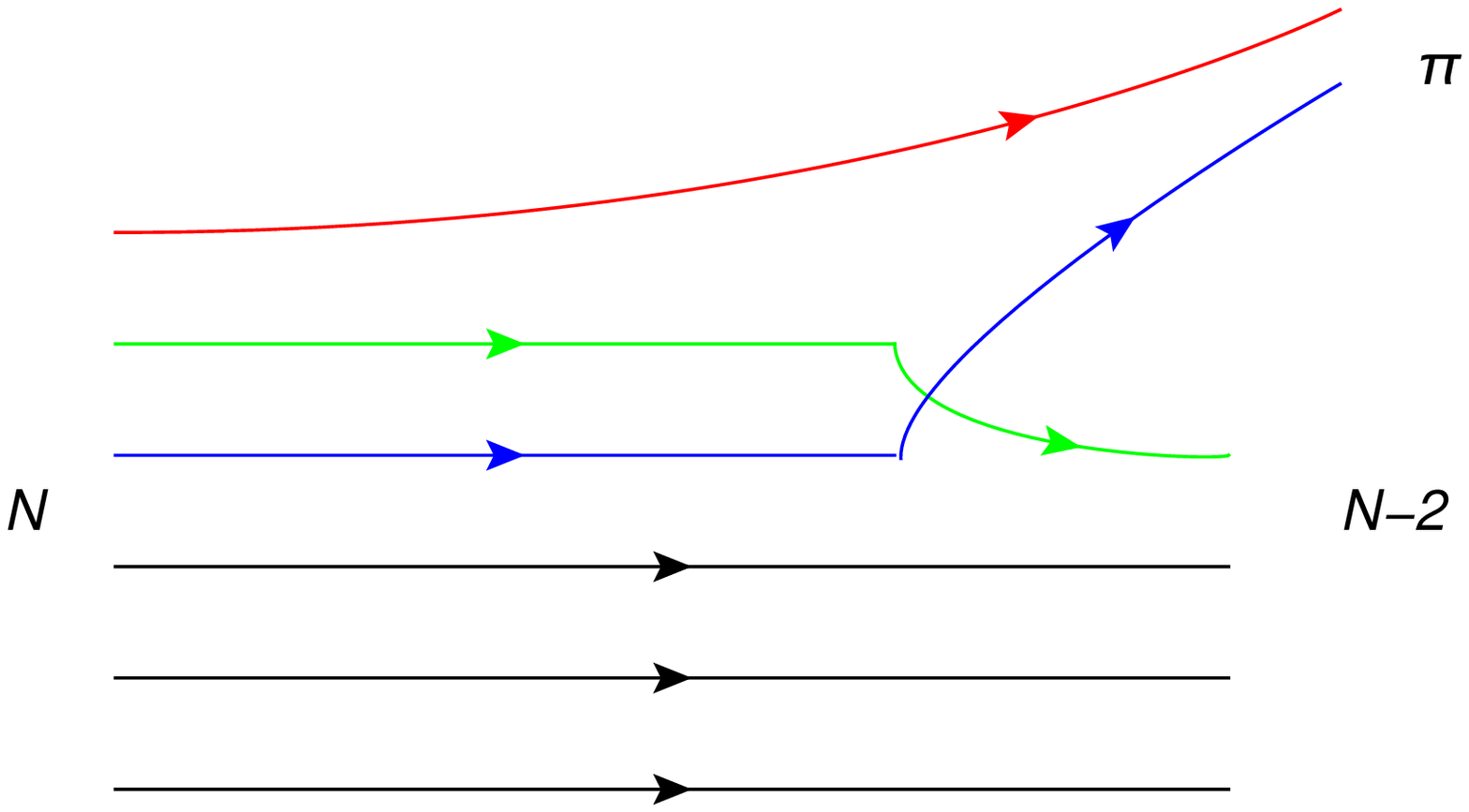}\:\:\raisebox{40pt}{+}\:\:	\includegraphics[scale=0.4]{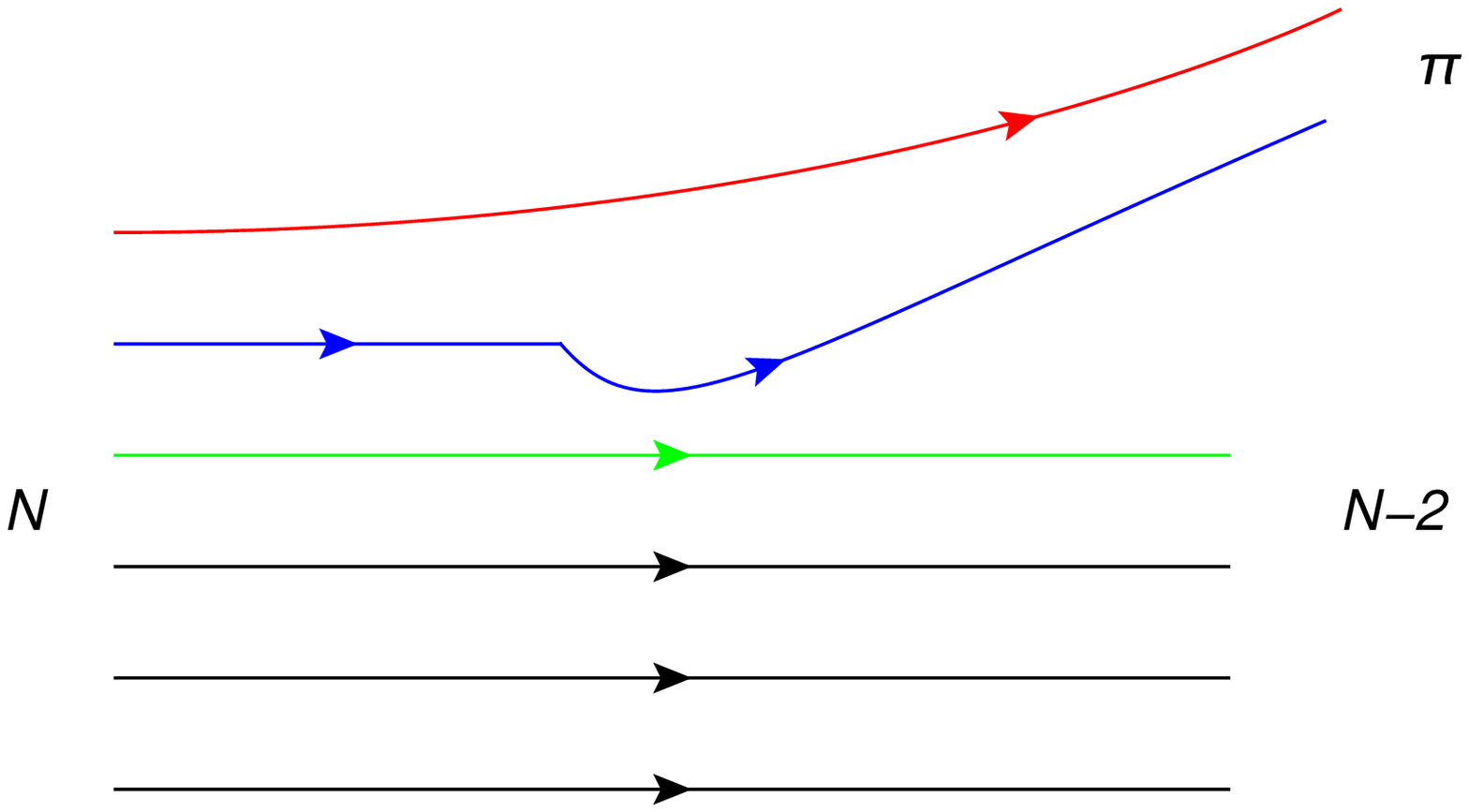} 
 		\label{fig2}
 \end{equation}
 The colored lines indicate the flow of $SO(2N)$ flavor among the particles participating in the process. The black lines denote the background fermions. The emitted pion carries away the flavor quantum numbers of two fermions. 
    
 In order to merge in a single pion, the off-shellness of the bound state fermions must be taken into account. This is done by considering the interaction of one of them with other background fermions.  The rate of the process can be easily estimated and to leading order in $1/N$ is given by, 
 \begin{equation} \label{evap}  
    \Gamma_{\pi} \sim m_f \alpha^2 \frac{1}{f_{\pi}^2} N^3  \sim 
        m_f \sim \frac{1}{R} \,.
 \end{equation}
Here the factor $\alpha^2 \sim 1/N^2$ comes from the contact  interaction among the fermions, the factor  $\frac{1}{f_{\pi}^2} \sim 1/N$ comes from the fermion-pion interaction, and, finally, the factor $N^3$ is the combinatoric one. 
  
 The above rate is similar  to the Hawking evaporation rate of a black hole of temperature 
  \begin{equation} \label{TTT}
  T \sim 1/R \sim m_f\,.  
  \end{equation}
 Notice that, just  as in the case of a black hole, the emission of more energetic quanta  (of energy $E \gg T \sim m_f$) is exponentially suppressed. This is because such emission requires a re-scattering of  large number of fermions into a single pion. Such processes are exponentially suppressed by a factor $e^{-E/m_f}$. Taking into account (\ref{TTT}), this gives a Boltzmann type suppression by $e^{-E/T}$.  Note, at initial stages of the decay, the deviations from thermality are of order $1/N$ ($1/S$), as this must also be the case for black holes \cite{Dvali:2012rt, Dvali:2015aja}.

 We must note that the origin of Hawking type radiation in the present case, as well as in case of a generic saturon, is very similar to the explanation of Hawking radiation in the picture of the black hole quantum $N$-portrait \cite{Dvali:2011aa}.  In this theory,  the black hole is described as a saturated state of soft gravitons. Their occupation number and coupling satisfy the relation $N = 1/\alpha_{gr}$. This is identical to the relation satisfied by the Gross-Neveu saturon. 

 The Hawking radiation in black hole $N$-portrait is a consequence of re-scattering of the constituent gravitons. This results in graviton emission, similar to the emission of a pion in the present model.  The two processes exhibit the identical scalings with $N$.  Of course, in case of a black hole, there is no specific need in introducing either the fermions or pions, since the theory already possesses black hole constituents in form of gravitons. 

However, the species composition is not important. Rather, the defining property for a black hole is that it represents a saturated state of gravitons \cite{Dvali:2011aa}.  This property controls  how a nearly thermal radiation can emerge from the decay of a pure state. Due to saturation, up to $1/N$ ($1/S$)-effects, the emission is thermal-like.

In the light of above, the presented case of Gross-Neveu saturons indirectly supports the picture of black hole compositeness \cite{Dvali:2011aa}. While in gravity one could be worried about the unknown effects, in Gross-Neveu model the same mechanism is calculable. 
 
The Gross-Neveu example explains why a saturated system, despite carrying a maximal microstate entropy, does not release the quantum information until the very late stages of its evaporation. Indeed, it is obvious that the  quantum information encoded in the flavor structure of the saturon, is slowly drained away by the emitted pion radiation.  However, this information is not resolvable at the initial stages of the decay, because of the following  two reasons. 
  
 First, each pion takes away only a $1/N$-fraction of the flavor content.  Based on this fraction, it is impossible to reconstruct the information pattern carried by the saturon. 
  
 Secondly, an observer has to pay an enormous probability price even for resolving the quantum number of a single outgoing pion. Since the pion coupling is $\sim 1/\sqrt{N}$, the minimal time-scale required for deciphering its quantum number,  even in a maximally efficient detector, is given by, 
  \begin{equation} \label{PT}  
    t_{\pi} \sim  N R  \,.  
 \end{equation}
 Therefore, in order to be able to start decoding any significant fraction of quantum information of the saturon, an observer has to collect at lest $\sim N$ pion quanta.   The time required for emitting this amount of quanta, is the same as (\ref{PT}). Not surprisingly,  this time-scale coincides with (\ref{TMIN}). 
 
 Thus, to conclude, there exists an absolute lower bound (\ref{TMIN}) on the time required for the start of  the information retrieval from the Gross-Neveu saturon.  This expression matches the universal formula (\ref{time}) for the information-retrieval time for a generic saturon, derived in \cite{Dvali:2020wqi}.  Black holes are no exception from this rule. The same time-scale (\ref{time}) in gravity coincides with Page's time. However, the Gross-Neveu model makes its origin very transparent.  This model shows that, as argued in \cite{Dvali:2020wqi}, the origin of time (\ref{time}) is not rooted in gravity but rather in the phenomenon of saturation.

\section*{Conclusions}

 In this paper we have tested the previously proposed \cite{Dvali:2020wqi}  (see also, \cite{Dvali:2019ulr,Dvali:2019jjw}) universal connection  between unitarity and entropy, in Gross-Neveu model. We found that the Gross-Neveu model has states with maximal entropy, namely $S_{max}=2N$, where $N$ is a number of Dirac fermion flavors in the model. Analysing the large-$N$ scaling of the $S$-matrix, we established a correlation between entropy and unitarity of scattering amplitudes. 

We observed that the entropy of the bound state saturates the unitarity bound (\ref{unba}) when the t'Hooft-like coupling is critical $\lambda \sim 1$. Upon saturation, the Gross-Neveu bound state becomes a saturon and assumes the properties similar to a black hole. In particular, the form of its entropy is identical to Bekenstein-Hawking entropy. An unstable saturon evaporates via Hawking's thermal rate with the temperature given by the inverse size. 
 
In addition, the minimal time-scale for the information retrieval (\ref{time}) is in one-to-one correspondence with Page's time. In the strict limit of infinite $N$, the information stored in the Gross-Neveu bound-state becomes unreadable. In order to retrieve it, one needs to resolve $1/N$ corrections. This is very similar to a situation with a black hole in its description as of bound-state of $N$ gravitons \cite{Dvali:2011aa}.  Our observations support the general points of \cite{Dvali:2020wqi} that black holes properties are not specific to gravity, but rather are the universal features of saturation.  \\
 
 {\bf Note added}
 
We must stress that black hole like properties of a saturon, including Hawking radiation,  where demonstrated recently in a parallel work \cite{Bubble}. In that paper, saturons represent the vacuum bubbles in a four dimensional theory with spontaneously broken $SU(N)$ symmetry, originally introduced in \cite{Dvali:2020wqi} as a model for saturation. Unlike the Gross-Neveu theory, the spectrum of bubbles is unbounded from above. However, the saturated bubble is the one with both its energy and entropy scaling as $\sim N$. It is shown in \cite{Bubble} that such bubbles behave as black holes. This is strikingly similar to the behaviour of the saturon of the Gross-Neveu theory discussed in the present article, despite the fact that theories are fundamentally different. The origin of this similarity is in the saturation of the bound (\ref{unba}), which again supports the point of \cite{Dvali:2020wqi} about universality of saturated states. It is desirable to come back to a more detailed discussion of how the correspondence between the states in the above two very different theories emerges at the saturation point of the entropy bound.

\section*{Acknowledgements}
This work was supported in part by the Humboldt Foundation under Humboldt Professorship Award, by the Deutsche Forschungsgemeinschaft (DFG, German Research Foundation) under Germany's Excellence Strategy - EXC-2111 - 390814868, and Germany's Excellence Strategy  under Excellence Cluster Origins.

\end{document}